 \documentclass[rog]{AGUTeX}

\usepackage{graphicx}
\usepackage{amssymb}
\usepackage{color}

\authorrunninghead{WANG}

\titlerunninghead{ Waves in Solar Coronal Loop}

\authoraddr{Corresponding author: Tongjiang Wang, USA. \\
(tongjiang.wang@nasa.gov)}

\begin{document}


\title{ \center Chapter 23: \\
  Waves in Solar Coronal Loops}


\authors{\center \large Tongjiang Wang\altaffilmark{1,2} }
\vspace{3mm}

\altaffiltext{1}{Department of Physics, The Catholic University of America,
   620 Michigan Avenue NE, Washington, DC 20064, USA; tongjiang.wang@nasa.gov, USA.  }

\altaffiltext{2}{NASA Goddard Space Flight Center, Code 671, Greenbelt, MD 20770, USA}

\begin{abstract}
Recent observations have revealed the ubiquitous presence of magnetohydrodynamic (MHD) waves and oscillations in the solar corona. The aim of this review is to present recent progress in the observational study of four types of wave (or oscillation) phenomena mainly occurring in active region coronal loops, including (i) flare-induced slow mode oscillations, (ii) fast kink mode oscillations, (iii) propagating slow magnetoacoustic waves, and (iv) ubiquitous propagating kink (Alfvenic) waves. This review not only comprehensively outlines various aspects of these waves and coronal seismology, but also highlights the topics that are newly emerging or hotly debated, thus can provide readers a useful guidance on further studies of their interested topics.
\end{abstract}

\begin{article}

\section{Introduction}
The corona is the outer part of the Sun's atmosphere with the high temperature as much as 
a few million kelvins (MKs), characterized by highly-structured and dynamic loops when observed 
in the X-ray band and in the extreme ultraviolet (EUV) bands. The corona is visible 
in the optical band only during a total solar eclipse or with a coronagraph. Coronal loops are 
believed to be plasma-filled closed magnetic flux anchored in the photosphere. 
Based on the temperature regime they are generally classified  into {\it cool} ($<$1 MK),
{\it warm} ($\sim$1.5 MK), and {\it hot} ($>$2 MK) loops \citep{rea14}. 
The magnetized coronal structures support propagation of various types of magnetohydrodynamics 
(MHD) waves. As such waves carry information about the structure and the physical properties 
of the medium, we can determine physical parameters of the corona that cannot be measured 
directly via a technique called {\it coronal seismology} \citep{uch70, rob84, nak05}. 
Knowledge of the coronal properties (e.g., magnetic fields, transport coefficients, 
inhomogeneous length scales) are crucial for enhancing our understanding of many fundamental 
but enigmatic processes, such as coronal heating \citep[see reviews by][]{kli06, klim14, par12}
and acceleration of the solar wind \citep[see reviews by][]{ofm10, cra12}.
 
The early evidence for coronal waves was mainly inferred from the periodicity observed 
in time profiles of fluctuations in radio and X-ray flux \citep[see reviews by][]{asc87, asc03}.
The successful launch of the Solar and Heliospheric Observatory
(SOHO) and the Transition Region and Coronal Explorer (TRACE) spacecrafts for the first time
enabled us to directly detect coronal wave activity with EUV imaging observations, 
which led to a variety of discoveries such as global EUV waves \citep{tho98}, 
flare-generated kink-mode loop oscillations \citep{asc99, nak99}, standing slow-mode loop
oscillations \citep{wan02, wan03a}, and propagating slow magnetoacoustic waves
in polar plumes \citep{ofm97, ofm99, def98}, and in coronal loops \citep{ber99, dem00}. 
Since then a great progress has been made 
in ground- and space-based imaging observations of coronal waves in almost all wavelengths. 
For example, propagating fast magnetoacoustic wave trains were observed along a coronal loop 
during a full solar eclipse \citep{wil01, wil02, kat03}, in post-flare supra-arcades
with TRACE \citep{ver05}, and along faint funnel-like coronal loops
recently discovered by the Atmospheric Imaging Assembly (AIA) onboard the Solar Dynamics 
Observatory (SDO) \citep{liu10, liu11}. A global fast sausage-mode wave was identified 
in radioheliograph observations of flaring loops \citep{asa01, nak03}. 
Ubiquitous propagating kink (Alfv\'{e}nic) waves in coronal loops were revealed with 
ground-based optical observations by the Coronal Multi-channel Polarimeter (CoMP)
\citep{tom07, tom09}. 

Stimulated by dramatically increasing imaging and spectroscopic observations, 
considerable progress has been also made in wide applications of coronal seismology 
over the last decade. Observational reviews on this subject can be found in 
\citet{asc03, asc04, asc12}, \citet{nak05}, \citet{dem05, dem08}, 
\citet{wang04, wang05, wang11}, \citet{nak07}, \citet{ban07}, \citet{dem12}, and
\citet{liu14}. Reviews of theoretical aspects can be found 
in \citet{rob00, rob02, rob04, rob08}, \citet{rob03}, \citet{goo05, goo06},
\citet{erd06}, \citet{ofm09}. In particular, two volumes of Space Science Reviews
\citep{nakerd09, erd11} and an IAU Symposium \citep{erdm08} were dedicated 
to detailed descriptions of various aspects of MHD waves and coronal seismology.

The aim of this review is to reflect recent progress in the observational study 
of four types of wave (or oscillation) phenomena mainly occurring in coronal loops
of active regions (ARs), including (i) flare-excited slow-mode waves, 
(ii) impulsively-excited kink-mode waves, (iii) propagating slow magnetoacoustic waves,
and (iv) ubiquitous propagating kink (Alfv\'{e}nic) waves. This review not only comprehensively 
outlines various aspects of these waves and coronal seismology, but also highlights
the topics that are newly emerging or hotly debated, thus can provide the reader 
a useful guidance on further studies. In addition, flare-induced
propagating fast-mode wave trains are a new wave phenomenon discovered in the corona 
with SDO/AIA \citep{liu10, liu11}, which was recently reviewed by \citet{liu14}, 
thus will not be included in this review.

\begin{table*}
\caption{Major types of MHD waves identified in coronal loops} \label{tab-waves}
\begin{tabular}{llllll}
\hline
\textbf{Wave type} & \textbf{Period} & \textbf{Phase speed\tablenotemark{a}} & \textbf{T$_e$} & \textbf{Observations} & \textbf{References }\\
   &  & (km/s) & (MK) & & \\
\hline
\underline{\textbf{Standing Waves}} & & & & \\
Slow mode & 7--31 min & 300--600  & 6--10  & SUMER Fe\,{\sc xix}/Fe\,{\sc xxi} & \citet{wan03b}  \\
          & 8--18 min & 300--400  & 6--10  & SUMER, Yohkoh/SXT & \citet{wan07}    \\
          & 3--8 min  & --        & 12--14 & Yohkoh/BCS & \citet{mari06} \\
          & 13--14 min & 300 &  7--18 & Radio 17 GHz, AIA 335 \AA & \citet{kim12}  \\
Kink mode & 2--11 min & 240--1660  & 1  & TRACE 171\AA & \citet{asc02}  \\
          & $\sim$16 min  & 600-1100 & 2 & NOGIS Fe\,{\sc xiv} & \citet{hor07}   \\
          & 2--10 min & 1000--3600  & 1  & SDO/AIA 171 \AA & \citet{whiv12}  \\
          & 12--80 min  & -- & 1-1.5  & SDO/AIA 171/193 \AA & \citet{liu14}  \\
          & 5 min & 3100  & 9--11 & SDO/AIA 131/94 \AA & \citet{whi12}  \\
Fast sausage mode & 14--17 s & 3200 & 5-10 & Nobeyama radio & \citet{nak03}  \\
                 & & & & & \citet{mel05}  \\
\hline
\underline{\textbf{Propagating Waves}} & & \textbf{Prop. speed} & & & \\
Slow-mode waves  & 10--15 min & 75--200 & 1.5 & SOHO/EIT 195 \AA & \citet{ber99} \\
           & 2--9 min & 70--235 & 1 & TRACE 171 \AA & \citet{dem00, dem02a} \\
           & $\sim$12 min & 132 & 1 & STEREO/EUVI 171 \AA & \citet{mar09}  \\
           & 12 \& 25 min & 100--120 & 1 & Hinode/EIS Fe\,{\sc xii} & \citet{wan09b} \\
           & 2--6 min & 30--150 & 0.4--1.2 & SDO/AIA 131/171/193 \AA & \citet{kid12}  \\
           & 10 \& 16 min & 70--100 & 1--1.5 & SDO/AIA 171/193 \AA & \citet{kri12b}\\
Reflected slow waves & -- & 160--350 & 2 & NOGIS Fe\,{\sc xiv} & \citet{hor07} \\
   & 11 min & 460--510 & 8--10 & SDO/AIA 131/94 \AA & \citet{kum13} \\
Fast-mode waves  & 6 s & 2100 & 2 & SECIS Fe\,{\sc xiv} & \citet{wil02} \\ 
                 & 90--220 s & 100--500 & 20 & TRACE 195 \AA (Fe\,{\sc xxiv}) & \citet{ver05} \\
                 & 25--400 s & 500--2200 & 1 & SDO/AIA 171 \AA & \citet{liu10, liu11} \\
        & & & & & \citet{liu14} \\
Kink (Alfv\'{e}nic) waves & $\sim$5 min & 600--700 & 1.8 & CoMP Fe~{\sc xiii} & \citet{tom07}  \\
 & & & & & \citet{tom09} \\
\hline
\end{tabular}
\tablenotetext{a}{The phase speed is estimated based on the fundamental mode.}
\end{table*}

\section{MHD Modes and Identification}
It is known that a magnetized homogeneous medium supports three basic MHD waves: 
the slow-mode wave, the fast-mode wave, and the incompressible Alfv\'{e}n wave. 
In the highly-structured solar corona, a magnetic cylinder is generally considered as 
an ideal model of common structures such as coronal loops.
Under usual coronal condition (i.e., $C_s<C_A<C_{Ae}$, where $C_s$ is the sound speed inside
the magnetic cylinder, $C_A$ and $C_{Ae}$ are the internal and external Alfv\'{e}n speeds), 
the linear MHD wave theory predicts an infinite variety of trapped modes in coronal loops 
depending on their radial wavenumber $l$, azimuthal wavenumber $m$, and longitudinal
wavenumber $n$ \citep{edw83, rob84}. However, generally only low order modes are detectable
\citep[for instance, the modes with $l$=0, $m$=0--1, and $n$=1--3; see a review by][]{nak07}.
The slow modes, fast kink modes, fast sausage modes, and torsional Alfv\'{e}n 
modes are four main types of MHD modes in a coronal loop.

  \begin{figure*}
  \noindent\includegraphics[width=40pc]{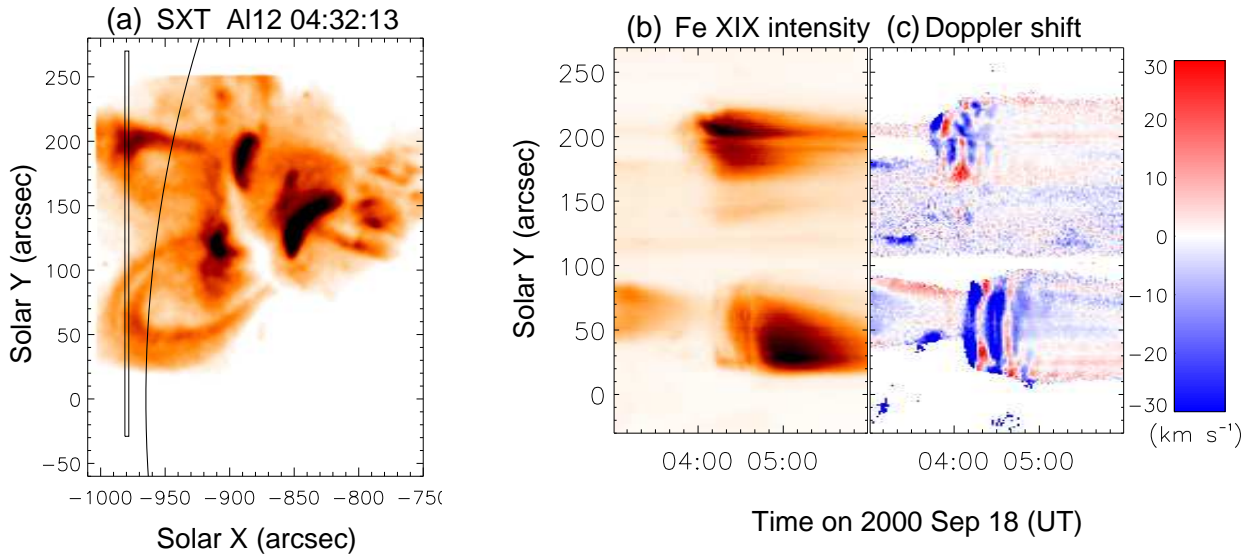}
  \caption{Hot loop oscillations observed by SOHO/SUMER \citep[from][]{wan07}. 
(a) Yohkoh/SXT image, indicating that the SUMER slit was located at the apex of 
two soft X-ray loops. Time distance plots of the Fe~{\sc xix} line intensity (b) and 
Doppler shift (c), showing that the fundamental standing slow-mode waves are detected 
as Doppler shift oscillations with a quick decay in the two loops. } 
  \label{fig1}
  \end{figure*}

The slow modes have a characteristic phase speed near the sound speed with only 
a weak dispersion. As the slow mode is dominated by longitudinal motions, it is often 
referred to as {\it the longitudinal mode}. The slow-mode waves propagate almost following 
magnetic field lines in the low-$\beta$ coronal condition. The kink modes are the branch of the
fast-mode regime when $m$=1, corresponding to asymmetric oscillations 
of a flux tube that appear as periodic transverse displacements. As the kink modes are dominated 
by the restoring force of magnetic tension with only weak compression, 
some studies also suggest to name it 
as {Alfv\'{e}nic} waves or {\it surface Alfv\'{e}n} waves \citep{goo09, goo12, mci11}. 
The kink modes have a phase speed lying between $C_A$ and $C_{Ae}$.
In the long-wavelength regime of $ka\ll$1 where $k=2\pi/\lambda$ is the wavenumber and 
$a$ the loop radius, the phase speed of the kink modes is equal to the kink speed, 
$C_k$, defined by $C_k=(2\rho_0/(\rho_0+\rho_e))^{1/2}C_A$, 
where $\rho_0$ and $\rho_e$ are the plasma densities inside and outside the tube. 
The fast sausage modes are the branch of the fast-mode regime when $m=0$,
corresponding to symmetric radial oscillations of a flux tube. The fast sausage
modes are strongly dispersive and have a cutoff at small wavenumbers, where the waves 
propagate with a speed close to $C_{Ae}$. 
If wavenumbers are too small, the waves become leaky \citep[called {\it leaky} 
modes, see][]{cal86}. Because of this cutoff, the trapped global (fundamental) sausage modes 
can exist only under special conditions 
\citep[e.g., in short and dense flare loops, see][]{nak03, asch04}. 
The detectable global sausage leaky modes may be
supported in long loops with realistic internal to external density contrast \citep{pas07}.
The last one, torsional Alfv\'{e}n modes, are axisymmetric with $m=0$, showing as periodic 
magnetic twist and plasma rotation \citep{van08a}.
As the torsional waves are completely incompressible, they can only be identified based
on Doppler shift patterns using spectroscopic observations \citep{wil04, dem12}.

Both {\it standing waves} with fixed spatial nodes (zero displacement) and {\it propagating
waves} are supported in coronal loops. Because a coronal loop has the natural node at
both endpoints due to the photospheric line-tied condition of the magnetic field, the standing
waves (also called eigen-modes) generally set up because of the wave reflection, 
while the propagating waves could exist if the wavelength is much small compare to 
the loop length, or they have insufficient time to develop a standing wave due to strong damping.
Table~1 lists the major types of MHD waves identified in coronal loops based on their 
characteristic propagation speeds (for propagating waves), oscillation periods 
(for standing waves),  as well as spatial features predicted by eigenfunctions. 
\vspace{2cm}

\section{Slow-Mode Oscillations of Hot Coronal Loops}
\subsection{Overview of Properties from SUMER Observations}
\vspace{5mm}

\subsubsection{Fundamental standing modes}
\subsubsection{Exciters}
\subsubsection{Decay}
\subsection{New Properties from SDO/AIA}

\section{Fast Kink-Mode Oscillations}
\subsection{Overview of flare-induced oscillations}
\subsection{Exciters}
\subsection{Decay and undecay}
\subsection{Horizontal and vertical polarizations}
\subsection{Multiple harmonics}
\subsection{Oscillating loops with flows}
\subsection{Persistent decayless oscillations}

\section{Propagating Slow-Mode Waves}
\subsection{Overview of properties for propagating coronal disturbances (PCDs)}
\subsection{Evidence of propagating slow waves in coronal structures above sunspots}
\subsection{Debate on interpretations of the PCDs in loops above plages}
\vspace{3mm}

\subsubsection{Properties of coronal outflows and PCDs}
\subsubsection{Towards resolving the controversy}

\section{Propagating Kink-Mode Waves}
\subsection{Observed properties}
\subsection{Theoretical interpretation}

\section{Final Remarks}

\footnote{\color{red} \normalsize Contact the author (email: tongjiang.wang@nasa.gov) to request an electronic version.}

\begin{acknowledgments}
This work was supported by NASA grants from NNX12AB34G and the NASA Cooperative Agreement
NNG11PL10A to CUA. I thank Pankaj Kumar and Steven Tomczyk for providing the original figures. 
Figures 1, 2, 4, 5 and 6 are reproduced by permission of the AAS. Figure 3 is reproduced
by permission of Astronomy \& Astrophysics, $@$~ESO.
\end{acknowledgments}


\end{article}

\end{document}